\documentclass[trackchanges, twocolumn]{aastex7}

\usepackage{xcolor}
\usepackage{colortbl}        
\usepackage{mathtools}
\usepackage{threeparttablex}
\usepackage{longtable}
\usepackage{booktabs}
\usepackage{rotating}
\usepackage{multirow}
\usepackage{soul}
\usepackage{subcaption}
\usepackage{url}
\usepackage{array}  
\usepackage{tabularx}
\usepackage{lipsum}
\usepackage{capt-of}
\newcolumntype{P}[1]{>{\centering\arraybackslash}p{#1}}
\usepackage[normalem]{ulem} 
\usepackage{graphicx}
\usepackage{ulem}
\usepackage{color}

\begin{document}

\title{Explaining the Origin of TeV Gamma Rays from M87 During High and Low States}

\author[orcid=0009-0009-8995-281X,gname=Nibedita,sname='Mondal']{Nibedita Mondal}
\affiliation{Astronomy \& Astrophysics Group, Raman Research Institute, Sadashivanagar, Bangalore 560080, Karnataka, India}
\email[show]{mon.nibe.mcc@gmail.com}

 \author[orcid=0000-0003-2445-9935,gname=Sandeep Kumar, sname='Mondal']{Sandeep Kumar Mondal} 
 \affiliation{Tsung-Dao Lee Institute, Shanghai Jiao Tong University, 1 Lisuo Road, Shanghai, 201210, People’s Republic of China}

 \email[show]{sandeep@sjtu.edu.cn}

 \author[orcid=0000-0002-1188-7503,gname=Nayantara,sname=Gupta]{Nayantara Gupta}
 \affiliation{Astronomy \& Astrophysics Group, Raman Research Institute, Sadashivanagar, Bangalore 560080, Karnataka, India}
 \email{nayan@rri.res.in}

\begin{abstract}
The detection of very high-energy gamma-rays from M87 can provide crucial insights into particle acceleration and radiation mechanisms in jets. The recent observations by the Large High Altitude Air Shower Observatory (LHAASO) detector extend the energy range of TeV gamma-ray astronomy, and also the variability study to the TeV energy domain. We have modelled the low state and flare state multi-wavelength spectral energy distributions of M87 within a time-dependent framework. In our model, the low state gamma-ray flux results from the emissions from the sub-parsec and the kilo-parsec scale jets of M87, whereas the flare state gamma-ray flux is mainly produced in the sub-parsec scale jet.  
We have shown that the spectral and temporal features of the TeV gamma-ray spectrum of M87 are consistent with this two-zone model, where the contribution from the sub-parsec scale jet significantly increases during the flare state.

\end{abstract}


\section{Introduction}
 
M87 is also known as 4FGL J1230.8 + 1223, NGC 4486, or 3C274, and commonly categorised as a Fanaroff-Riley I radio galaxy \citep{1974MNRAS.167P..31F} located at redshift z = 0.00428 \citep{10.1111/j.1365-2966.2010.18174.x} with Right Ascension (R.A) = 187.7059° and Declination (Decl.) = 12.3911°. 
\par
It is well known as one of the brightest radio sources in the sky since the beginning of radio astronomy \citep{Bolton1949}. The large-scale radio structure of this galaxy extends up to 80 kpc \citep{1952Natur.170.1063M}. The recent review paper by \cite{hada2024m} gives an excellent overview of different aspects of this source.

\par
It is one of the nearest radio galaxies, and has been identified as a VHE gamma-ray emitter in the TeV energy range. Since M87 is the only source that has been traced from the supermassive black hole (SMBH) shadow at about 0.005 pc \citep{2019ApJ...875L...1E, 2019ApJ...875L...2E, 2019ApJ...875L...3E, 2019ApJ...875L...4E, 2019ApJ...875L...5E, 2019ApJ...875L...6E} to the outer jet extending to $\sim$25\, kpc, exhibiting a relativistic jet, it has been considered to be the best-researched laboratory for studying SMBH physics, accretion, and radio-loud AGN properties.
\par

M87 exhibits brightness across the entire electromagnetic spectrum, which makes it an excellent candidate for broadband multi-wavelength observations. Multi-wavelength studies at different epochs reported emission across time scales from days to years, but due to insufficient data in different time domains, the underlying physics of temporal variations remains unclear \citep{hada2024m}. 
\par
Very-high-energy (VHE, $0.1\,\sim\,100~\mathrm{TeV}$) gamma-rays are messengers of cosmic ray acceleration sites. The origin of this VHE emission of jetted AGNs like M87 is still elusive. The High Energy Gamma Ray Astronomy (HEGRA) reported the first VHE gamma-ray emission from M87 in 1998–1999 \citep{aharonian2003giant}. Furthermore, the next-generation high-energy instruments such as High Energy Stereoscopic System (H.E.S.S.; \cite{2006Sci...314.1424A}), Very Energetic Radiation Imaging Telescope Array System (VERITAS; \cite{Acciari_2008}), Major Atmospheric Gamma Imaging Cherenkov (MAGIC; \cite{Albert_2008}), and \textit{Fermi} Large Area Telescope (\textit{Fermi}-LAT; \cite{Abdo_2009}) also reported gamma ray emission. Additionally, High Altitude Water Cherenkov Observatory (HAWC) recorded long-term VHE emission from this source at a weak statistical significance (3.6$\sigma$) \citep{Albert_2021}. According to previous studies \citep{2006Sci...314.1424A, Albert_2008, aliu2012veritas}, M87 not only exhibits variations in its spectrum and flux in flaring states but also displays complex behaviour in non-flaring states at gamma-ray frequencies \citep{2019A&A...623A...2A}. The following is a brief history of M87 in different activity states.

Four significant VHE flares have been detected in 2005, 2008, 2010, and 2018 \citep{2006Sci...314.1424A, Albert_2008, abramowski20122010, algaba2024broadband}, exhibiting one-day variability. LHAASO detected VHE flux state \citep{LHAASO_M87:2024} in 2022 for 8 days. Apart from these, the source has been seen in the quiescent state, with a high flux state occasionally appearing for very short time scales.

 During long-term monitoring from 2021 to 2024, LHAASO observed M87 with a statistical significance of $\sim9\sigma$, and the observed energy spectrum extended up to $20\,\mathrm{TeV}$ \citep{LHAASO_M87:2024}. The HAWC reported the emission from M87 over approximately 7.5 years (from 2015 to 2022) in the energy range of 16 to 26 TeV \citep{alfaro2025longtime}.

In this work, we have studied the jet emission from M87 with data analysis and modelling of the spectral energy distributions to address the key aspects of jet physics, in particular, the emission mechanisms and regions in the high-energy regime, in light of the gamma-ray data recently reported by the LHAASO observatory. 

The \textit{Fermi}-LAT data analysis has been discussed, and the possibility of a GeV flare in \textit{Fermi}-LAT data has been explained in section \ref{sec:MW_Data_Accumulation}. Subsequently, the Swift XRT and UVOT data analysis have been discussed, and the LHAASO data, the other archival data used in this study, have been mentioned in this section. Section \ref{sec:MWSED_modelling} is on the modelling of the spectral energy distribution covering the multi-wavelength observations, and our results are mentioned in section \ref{sec:result}. Discussion and conclusion are presented in section \ref{sec:discussion}.

\section{Multi-Wavelength Data ACCUMULATION} \label{sec:MW_Data_Accumulation}


\subsection{\textit{Fermi}-LAT Data Analysis} \label{subsubsec:Fermi_LAT_Data_Analysis}

The \textit{Fermi} Gamma-ray Space Telescope (FGST) is a space-based high-energy gamma-ray observatory. It carries two instruments: the Large Area Telescope (LAT), which serves as the primary instrument, and the Gamma-ray Burst Monitor (GBM). The LAT is an imaging, pair-conversion, wide-field-of-view telescope which can detect gamma-rays in the energy range from 20 MeV to beyond 300 GeV, with a field of view of 2.7 sr at 1 GeV and above \citep{Atwood_2009}. It scans the entire sky every three hours and has been in continuous operation since August 2008.
\par

For our work, we conducted a comprehensive analysis of \textit{Fermi}-LAT gamma-ray Pass 8 data for M87, using data collected from the Fermi Science Support Center (FSSC) server \citep{fermi_lat_query}, covering more than 16.5 years (4\textsuperscript{th} August 2008 to 6\textsuperscript{th} February 2025). The data were extracted from a circular region of a radius 30$^\circ$, centered on M87, covering the energy range 0.1–100 GeV.
\par    

The dataset was processed using Fermipy (v1.0.1; \cite{Fermipy_Version}), an open-source Python package for \textit{Fermi}-LAT data analysis. A 15$^\circ \times$15$^\circ$ region was defined in the configuration file, and only photons with energies between 0.1 GeV and 100 GeV were included, while events with zenith angles above 90$^\circ$ were excluded to reduce contamination from the Earth’s limb. For event selection, the ‘P8R3 SOURCE’ class (evclass=128) was used, as recommended for analyses with relatively small regions of interest (ROIs) less than 25$^\circ$ \citep{bruel2018fermi}. The evtype parameter was set to 3 to include all event types, covering both the front and back sections of the tracker. Data quality cuts were applied using
$ \texttt{DATA\_QUAL} > 0 \ \&\& \ \texttt{LAT\_CONFIG} == 1 $
to ensure that only high-quality data collected under standard LAT science operations were included.
\par
 
 For source modelling, the \textit{Fermi}-LAT Fourth Source Catalogue Data Release 4 (4FGL-DR4; gll\_psc\_v35.fits; \cite{4FGL_DR4}) was included, with the Galactic diffuse emission modelled using the gll\_iem\_v07 template \citep{Fermi-LAT_galactic_diffuse_model} and the isotropic extragalactic background modelled with iso\_p8r3\_source\_v3\_v1.txt. Within the region of interest (ROI), the parameters of 4FGL J1230.8+1223, together with the isotropic (isodiff) and Galactic diffuse (galdiff) components, were left free. 
 The normalisation (norm) of the diffuse components was allowed to vary, while for 4FGL J1230.8+1223, represented by a log-parabola, the norm, spectral index (alpha), and curvature (beta)were freed. The fit was performed using \texttt{gta.fit()} with the NEWMINUIT optimiser, iterating until the fit quality reached 3 to obtain the best-fit model.
 
\par 
The \textit{Fermi}-LAT data analysis has two objectives: (i) gamma-ray SED analysis and (ii) light curve analysis to investigate a potential GeV flare coinciding with the VHE flare from M87 in January 2022. 

LHAASO monitored M87 from MJD 59281-60385 \citep{LHAASO_M87:2024}, the details are given in Table~\ref{tab: Phase_Period}. Considering this timeline, we have analysed \textit{Fermi}-LAT gamma-ray spectral energy distribution (SED) for three states (low-state 1, flare state, low-state 2), using \texttt{gta.sed()} in Fermipy. 

\begin{deluxetable*}{l!{\vrule width 1.2pt}l}
\tablecaption{Time intervals of different states of M87 between 2021-2024 ( MJD 59281-60385), based on LHAASO observation \cite{LHAASO_M87:2024}   \label{tab: Phase_Period}} 
\tablewidth{0pt}
\tablehead{
  \colhead{\textbf{Time Interval (MJD)}} & \colhead{\textbf{State}}
}
\startdata
59281-59607 & Preflare (low-state 1 or LS1) \\
59607-59615 & Flare State\\
59615-60385 & Postflare (low-state 2 or LS2) \\
\enddata
\end{deluxetable*}

Following Fermipy documentation \citep{fermipy_docs}, the \textit{Fermi}-LAT gamma-ray light curve of M87 was then extracted in different time bins, as described in Sec :~\ref {sec:Fermi-LAT_Lightcurve}.

 \subsubsection{Gamma-ray : Likelihood Test Using Different Spectral Model}
 \label{sec:Fermi-LAT_mwsed}
 
 \textit{Fermi}-LAT data analysis incorporates the Maximum Likelihood (ML) method to determine the best-fit model of the SED data. We considered two spectral models : (i) PowerLaw (PL), and (ii) LogParabola (LP) for comparison\footnote{\url{https://fermi.gsfc.nasa.gov/ssc/data/analysis/scitools/source_models.html}}.

\begin{itemize}
    \item PowerLaw (PL) :
                \begin{equation}
	               \frac{dN}{dE}=N_{\mathrm{PL}}\Bigg(\frac{E}{E_{\circ}}\Bigg)^{-\alpha} 
	          \end{equation}
where, N$_{\mathrm{PL}}$ is the normalization factor, E$_{\circ}$ is the scaling factor or \verb|pivot energy|, and $\alpha$ is the spectral index of the PowerLaw (PL) model.  N$_{\mathrm{PL}}$ and $\alpha$ were kept free during this model fitting. The 4FGL catalogue assigns a \verb|pivot energy| E$_{\circ}$ to each source, which is the point at which the fit parameters have the least correlation and the spectrum is well measured. In our analysis, the scaling factor or pivot energy is fixed at its catalogue value of 1.14 GeV \citep{4FGL_DR4} for both the PowerLaw and LogParabola spectral models.

    \item LogParabola (LP): The LogParabola (LP) model is widely used in Gamma-ray astronomy, accounts for spectrum curvature, and is parameterised as follows:
\begin{equation}
	\frac{dN}{dE}=N_{\mathrm{LP}}\Bigg(\frac{E}{E_{\circ}}\Bigg)^{-\alpha+\beta log(E/E_{\circ}))} 
	\end{equation}
where, N$_{\mathrm{LP}}$ is the prefactor, the photon index $\alpha$, $\beta$ is the curvature index. N$_{\mathrm{LP}}$, $\alpha$, $\beta$ are considered free parameters during fitting. 
\end{itemize}
 
 Table~\ref{Tab:MLE_GammaR_SED_Param} shows the best-fit parameters from the likelihood analysis, where the LogParabola (LP) model best describes the gamma-ray spectrum in the flare and low states. 

\begin{table*}
\centering
\caption{Results of \textit{Fermi}-LAT SEDs of different states, fitted with different spectral models: PL, LP}
\label{Tab:MLE_GammaR_SED_Param}
\begin{tabular*}{\textwidth}{@{\extracolsep{\fill}}||c c c c c c c c c c||}
\hline
& & PowerLaw (PL) & & & & & & & \\
\hline
State & F$_{0.1 - 100 \ \mathrm{GeV}}$ & Index ($\Gamma$) &  &  & TS &  & -log(Likelihood) &  & \\ 
& ($10^{-8} \ \mathrm{ph} \ \mathrm{cm}^{-2} \ \mathrm{s}^{-1}$) & & & & & & & & \\ 
\hline\hline
Pre-Flare (LS1) & 2.028$\pm$0.009 & 2.080$\pm$0.002 &  &  & 158.06 &  & 102802.22 &  &  \\
\hline
Flare & 3.02$\pm$1.46 & 2.23$\pm$0.22 &  &  & 2.55 &  & 5278.21 &  &  \\
\hline
Post-Flare (LS2) & 1.36$\pm$0.21 & 1.95$\pm$0.07 &  &  & 307.11 &  & 176206.73 &  &  \\
\hline\hline
\end{tabular*}

\vspace{0.6cm}

\begin{tabular*}{\textwidth}{@{\extracolsep{\fill}}||c c c c c c c||}
\hline
& & & LogParabola (LP) & & & \\
\hline
State & F$_{0.1 - 100 \ \mathrm{GeV}}$ & $\alpha$ & $\beta$ & TS & -log(Likelihood) & $\Delta \log$(Likelihood) \\
& ($10^{-8} \ \mathrm{ph} \ \mathrm{cm}^{-2} \ \mathrm{s}^{-1}$) & & & & & \\
\hline\hline
\textit{Pre-Flare (LS1)} & \textit{1.723$\pm$0.001} & \textit{2.026$\pm$0.003} & \textit{0.073$\pm$0.001} & \textit{159.92} & \textit{102801.82} & \textit{-0.40 }\\
\hline
\textit{Flare} & \textit{1.39$\pm$0.42} & \textit{3.33$\pm$0.34} & \textit{1.84$\pm$0.24} & \textit{5.84} & \textit{5276.69} & \textit{-1.52} \\
\hline
\textit{Post-Flare (LS2)} & \textit{0.94$\pm$0.23} & \textit{1.79$\pm$0.12} & \textit{0.14$\pm$0.06} & \textit{312.38} & \textit{176203.38} & \textit{-3.35} \\
\hline\hline
\end{tabular*}

\tablecomments{Model preference is determined using the log-likelihood ratio test, where 
$\Delta\log(\mathrm{Likelihood}) = \log(\mathrm{Likelihood})_{\mathrm{LP}} - \log(\mathrm{Likelihood})_{\mathrm{PL}}$. 
The tabulated values appear negative because they are derived from $-\log(\mathrm{Likelihood})$, 
a smaller $-\log(\mathrm{Likelihood})$ (or equivalently, a larger $\log(\mathrm{Likelihood})$) corresponds to a better fit. 
Therefore, the negative $\Delta\log(\mathrm{Likelihood})$ values indicate that the LogParabola (LP) model yields higher likelihoods 
than the PowerLaw (PL) model and thus provides the statistically preferred fit in all activity states. 
The  $\Delta\log(\mathrm{Likelihood})$ values of the best-fitted models are highlighted in \textit{italic} font.}
\end{table*}
 
\subsubsection{Gamma-ray Spectral Analysis Outcomes}
\begin{itemize}
    \item Pre-flare State (low-state 1 or LS1): The source (4FGL J1230.8 + 1223) is detected with statistical significance TS=159.92 from our likelihood analysis. The Gamma-ray spectrum for Pre-flare state is well represented by a log-parabola model with $\alpha=2.026\pm0.003$, $\beta=0.073\pm0.001$ and corresponding integrated flux over 0.1-100 GeV of 1.723$(\pm0.001)\times 10^{-8}$ ph cm$^{-2}$ s$^{-1}$.
\end{itemize} 
\begin{itemize}
    \item Flare State: In this state, the Gamma-ray spectrum is defined by a log-parabola with $\alpha=3.33\pm0.34$, $\beta=1.84\pm0.24$ detected with TS=5.84. An integrated flux of 1.39 ($\pm$ 0.42)$\times 10^{-8}$ ph cm$^{-2}$ s$^{-1}$ over 0.1-100 GeV is obtained.
 
\end{itemize} 
\begin{itemize}
    \item Post-flare State (low-state 2 or LS2):
 The log-likelihood test for post-flare state signifies that a log-parabola model is a better representation over a single power-law model with an integrated flux of 0.94 ($\pm$0.23) $\times$ 10$^{-8}$ ph cm$^{-2}$ s$^{-1}$ over the energy range 0.1-100 GeV with best-fit indices $\alpha=1.79\pm0.12$ and $\beta=0.14\pm0.06$. The source is detected with high statistical significance TS=312.38, compared to other states.
\end{itemize} 
\par

\begin{figure*}[h]
    \centering

    \begin{subfigure}{0.9\linewidth}
        \centering
        \includegraphics[width=0.7\linewidth]{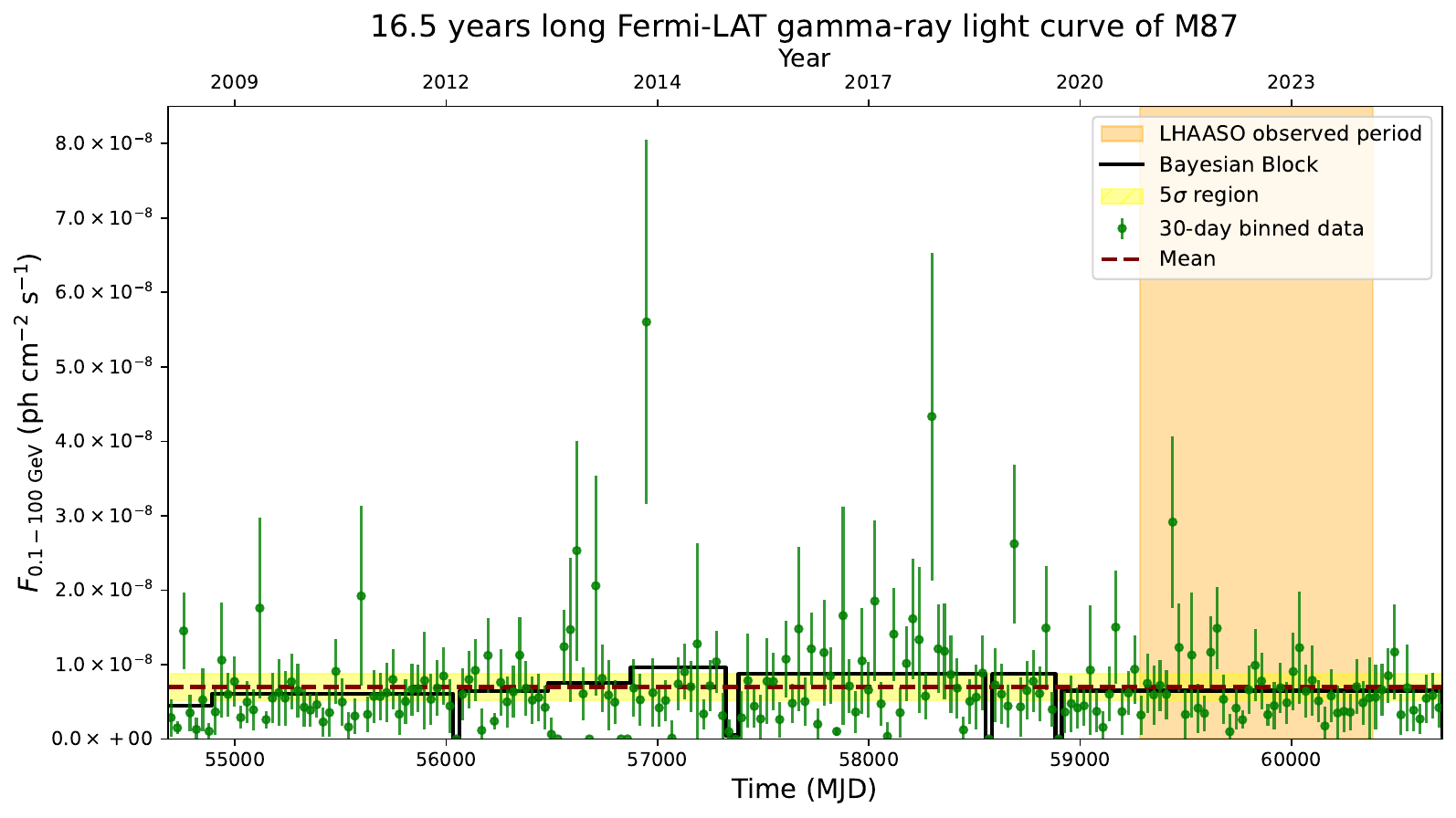}
        \caption{16.5-year (MJD 54682–60712) long 30-day binned light curve of M87. Orange shaded region: period of continuous LHAASO monitoring \citep{LHAASO_M87:2024}.}
        \label{fig:30D_Binned_Fermi_LAT_LC_16.5yrs}
    \end{subfigure}

    \begin{subfigure}{0.48\linewidth}
        \centering
        \includegraphics[width=\linewidth]{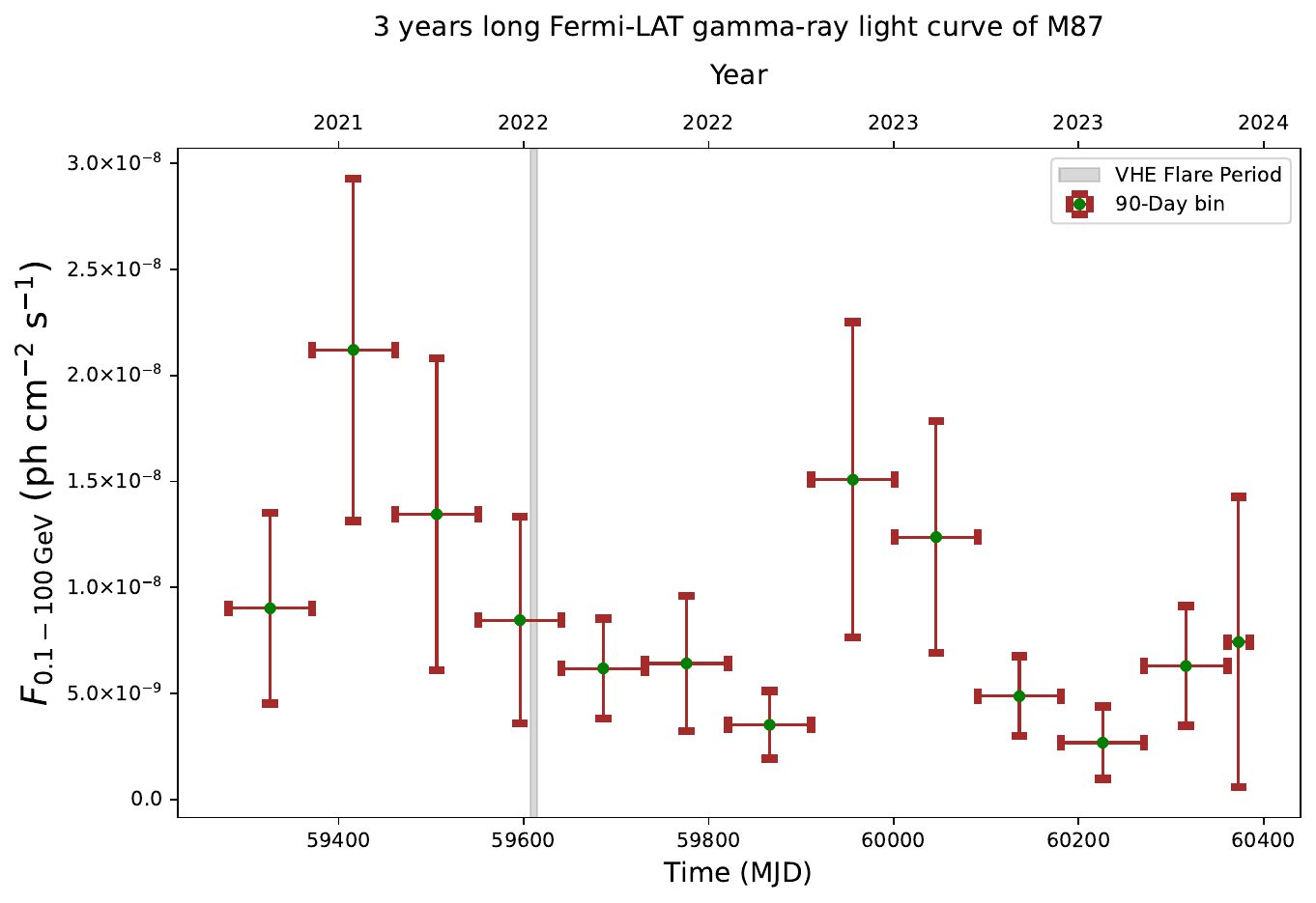}
        \caption{3-year (MJD 59281–60385) 90-day binned light curve.}
        \label{fig:3yr_90D_bin}
    \end{subfigure}
    \hfill
    \begin{subfigure}{0.48\linewidth}
        \centering
        \includegraphics[width=\linewidth]{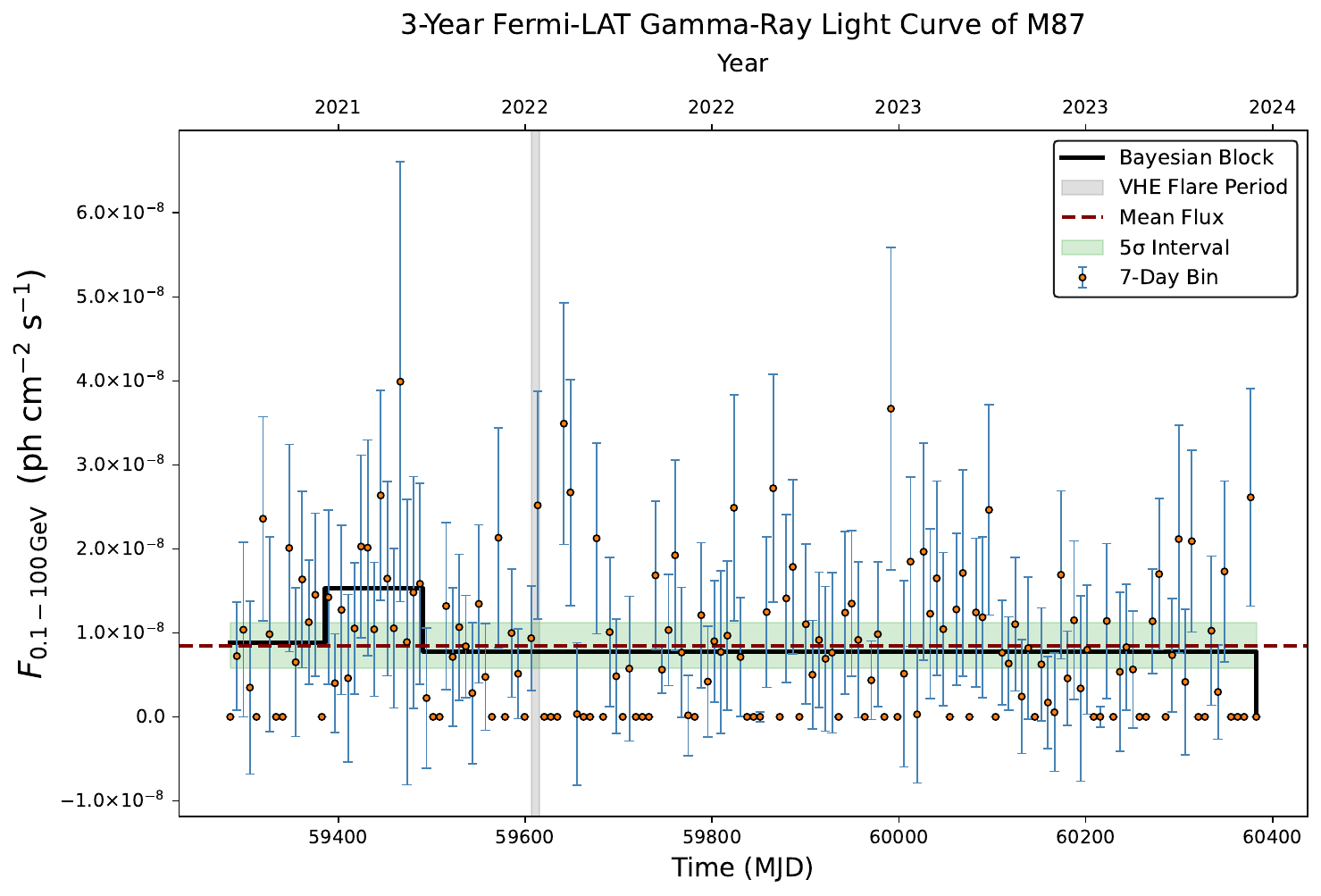}
        \caption{3-year (MJD 59281–60385) 7-day binned light curve.}
        \label{fig:7D_3Y}
    \end{subfigure}

    \begin{subfigure}{0.9\linewidth}
        \centering
        \includegraphics[width=0.7\linewidth]{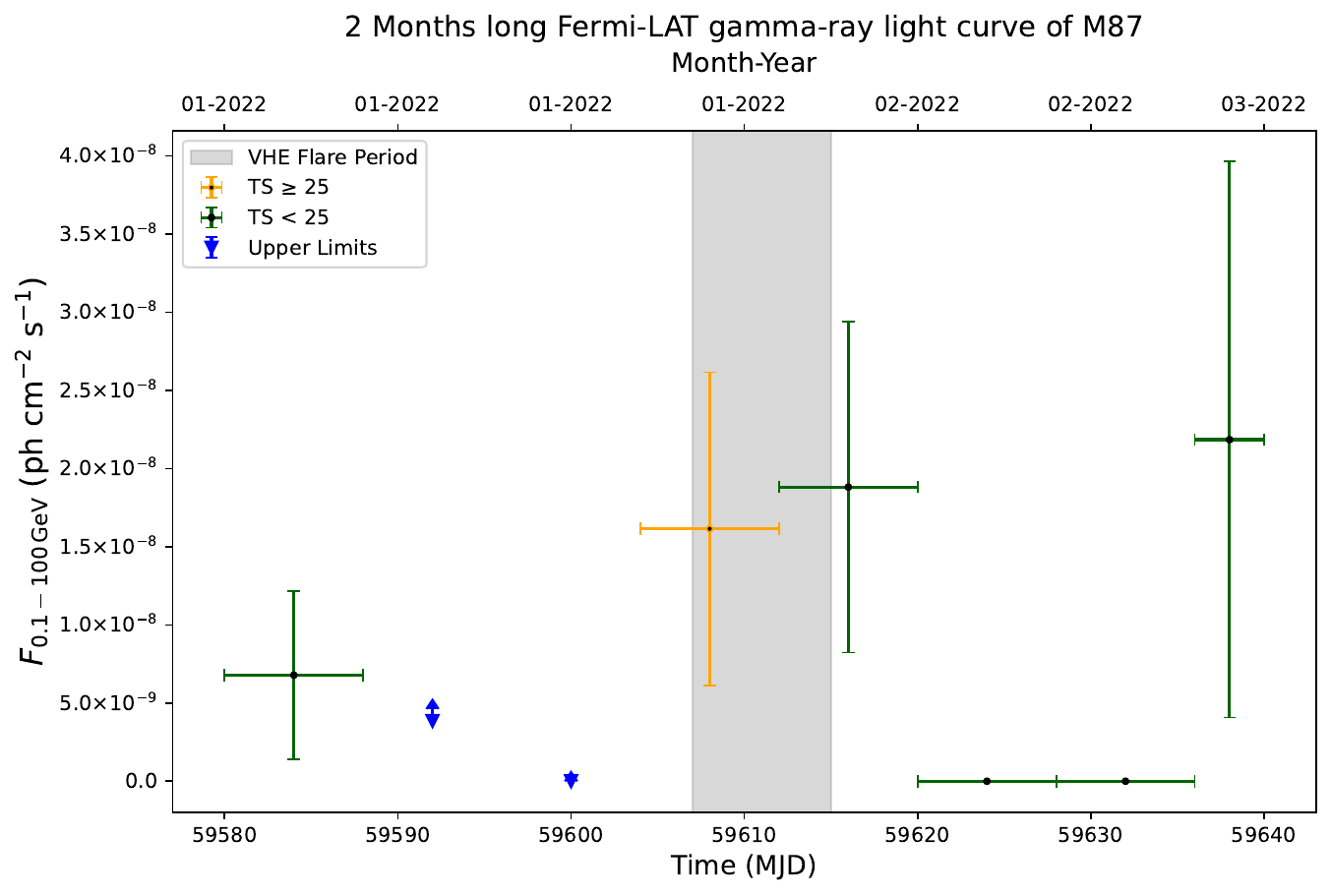}
        \caption{2-month-long 8-day binned light curve of M87.}
        \label{fig:8D_2M}
    \end{subfigure}

    \caption{Gamma-ray light curves of M87 observed with \textit{Fermi}-LAT at different binning at time periods.}
    \label{fig:M87_lightcurves}
\end{figure*}


\subsubsection{GeV flare in \textit{Fermi}-LAT} 
\label{sec:Fermi-LAT_Lightcurve}

We examined the possibility of a GeV flare in \textit{Fermi}-LAT coincident with the VHE flaring period detected by LHAASO in 2022 \citep{LHAASO_M87:2024} by analysing the \textit{Fermi}-LAT gamma-ray light curve over 16.5 years, from 4\textsuperscript{th} August 2008 to 6\textsuperscript{th} February 2025 (MJD 54682.65–60712.00) using 30-day bins. The Bayesian block method \citep{Bayesian_Block_2013} was applied to identify flux enhancements (Fig.~\ref{fig:30D_Binned_Fermi_LAT_LC_16.5yrs}). A horizontal maroon colour dashed line indicates the average gamma-ray flux over the full period,  while the Bayesian blocks are shown as a black solid line. The orange-shaded region highlights the interval during which LHAASO continuously monitored the source (MJD 59281–60385; \cite{LHAASO_M87:2024}).
\par
Statistically, no flux-enhanced state was detected during the LHAASO monitored period, i.e., MJD 59281–60385. There is a possibility that the flux may have increased slightly compared to the average gamma-ray flux between 2014 and 2019; however, this period falls outside the focus of our study. 
\par
Typically, after identifying flux enhancements in a light curve with larger time bins, we analyse the light curve with shorter binning to reveal possible sub-structures more clearly \citep{Mondal:2021vbs}. In this case, we followed \cite{LHAASO_M87:2024} and further analysed this 3-year \textit{Fermi}-LAT data (MJD 59281–60385) in 90-day bins (Fig.\ref{fig:3yr_90D_bin}), marking the LHAASO-detected VHE flare with a grey-shaded region. From Fig.\ref{fig:3yr_90D_bin}, it can be seen that the source is not in a high-flux state in the MeV–GeV band and the 90-day binned data have large flux uncertainties. These errors would increase further if we analyse the light curve in smaller time binning. To demonstrate this, we analysed the same \textit{Fermi}-LAT light curve using 7-day binning (Fig.~\ref{fig:7D_3Y}). The flux variation appears highly random, and the flux errors are very large. Using the Bayesian block method, we did not find any GeV flaring during the VHE flare (grey-shaded region), and the data quality is not feasible for further analysis in smaller time bins \citep{Mondal:2022zvv}. Finally, we analysed the $\sim$60-day light curve in 8-day bins (Fig.\ref{fig:8D_2M}), as done in \cite{LHAASO_M87:2024} also. Only one data point falls within the VHE flare period (grey-shaded region), with TS$>$25. Because only a single data point is available for this period and flux uncertainties increase when using smaller time bins, it is not possible to reliably determine whether \textit{Fermi}-LAT detected a flare coincident with the VHE event observed by LHAASO in 2022. Given this limitation, any claim of a flare state would be statistically unsupported. Reduced flux uncertainties would have allowed finer temporal binning and a more detailed examination of the light curve for potential flux enhancements. Therefore, the current \textit{Fermi}-LAT data does not provide statistically significant evidence for a GeV flaring event coincident with the VHE flare of M87.

\subsection{Swift XRT and UVOT Data Analysis}
\label{subsubsec:Swiftxrt_UVOTT_Data_Analysis}

The Neil Gehrels Swift is a multi-wavelength space-based observatory with
three instruments onboard: Burst Alert Telescope (BAT; 15 - 150 keV), X-Ray Telescope (XRT; 0.3 - 10 keV), and UVOT (170 - 600 nm) \citep{SWIFT:2005}. It observes the sky in hard X-ray, soft X-ray, Ultraviolet, and Optical wavebands. Swift observations of M87 were retrieved from the Swift Master Catalog (swiftmastr) using the HEASARC Xamin interface\footnote{\label{fnote:note4}\url{https://heasarc.gsfc.nasa.gov/xamin/}}
. We searched for observations centered on M87 by querying the catalog with the source name “M87” (J2000 position: RA = 187.7059$^\circ$, Dec = +12.3911$^\circ$). The default Xamin search parameters were used, corresponding to a circular search radius of 0.1$^\circ$. We found 54 Swit-XRT/ UVOT observations for the period MJD 59281 and 60385~\footref{fnote:note4}. Observations that yielded statistically insignificant data were excluded from the analysis. The discarded ObsIDs are listed in Table~\ref{tab:SWIFT_OBSlog}. No observations were found corresponding to the 8-day VHE flare (MJD 59607–59615), i.e., all the Swift observations correspond to LS1 \& LS2.

\begin{ThreePartTable}
\begin{TableNotes}[para,flushleft]
\footnotesize
\vspace{0.5em}
\item[*] Excluded only from Swift-XRT analysis.\\
\item[$\dagger$] Excluded from both Swift-XRT and UVOT analyses.\\
\vspace{0.5em}
\end{TableNotes}

\begin{longtable}{ | c |P{3cm}|P{3.5cm}|P{2cm}|P{2cm}|P{2cm}|} 
\caption{Discarded Swift-XRT and UVOT observations of M87 } 
 \label{tab:SWIFT_OBSlog}\\
 \insertTableNotes \\ 

\hline
\textbf{Sr. No.} & \textbf{Observation ID} & \textbf{Instrument} & \textbf{Starting Time (MJD)} & \textbf{XRT Exposure (ks)} & \textbf{UVOT Exposure (ks)} \\
\hline
\endfirsthead

\hline
 \textbf{Sr. No.} & \textbf{Observation ID} & \textbf{Instrument} & \textbf{Starting Time (MJD)} & \textbf{XRT Exposure (ks)} & \textbf{UVOT Exposure (ks)} \\
\hline
\endhead

\hline
\endfoot

\hline
\endlastfoot
 1 & 00031105075* & SWIFT-XRT/UVOT & 59307.253 & 0.9 & 0.9 \\
 2 & 00031105076* & SWIFT-XRT/UVOT & 59310.239 & 0.9 & 0.9 \\
 3 & 00031105078* & SWIFT-XRT/UVOT & 59315.143 & 0.6 & 0.6 \\
 4 & 00031105083\(^{\dagger}\) & SWIFT-XRT/UVOT & 59310.005 & 0.01 & 0.0 \\
 5 & 00031105093* & SWIFT-XRT/UVOT & 60048.201 & 0.8 & 0.8 \\
 6 & 00031105094*  & SWIFT-XRT/UVOT & 60051.318 & 0.9 & 0.9 \\
7 & 00096453001*  & SWIFT-XRT/UVOT & 59650.352 & 1.0 & 1.0 \\

8 & 00096453003* & SWIFT-XRT/UVOT & 59652.167 & 0.7 & 0.7 \\
9 & 00096453008*  & SWIFT-XRT/UVOT & 59658.257  & 0.6 & 0.6 \\
10 & 00096453009*  & SWIFT-XRT/UVOT & 59659.443  & 0.7 & 0.7 \\

 11 & 00096453011*  & SWIFT-XRT/UVOT & 59661.234 & 0.8 & 0.8 \\
 12 & 00096453012* & SWIFT-XRT/UVOT & 59662.498 & 0.9 & 0.9 \\
13 & 00096453013*  & SWIFT-XRT/UVOT & 59663.178 & 0.6 & 0.6 \\
14 & 00096453014* & SWIFT-XRT/UVOT & 59664.700 & 0.9 & 0.9 \\
15 & 00096453016*  & SWIFT-XRT/UVOT & 59666.410 & 0.7 & 0.7 \\

16 & 00096453028* & SWIFT-XRT/UVOT & 59677.353 & 0.7 & 0.7 \\

\end{longtable}
\end{ThreePartTable}

\begin{center}
    \begin{table}
    \centering
     \caption{Swift-XRT model parameter values for Pre-Flare (LS1) \& Post-Flare (LS2)}
        
    \label{tab:Swift_XRT_Fitting_Parameter}
    \begin{tablenotes}
      \small
	 \item *  denotes the parameter is fixed.
      \item $\dagger$ for using the updated APEC model.
     \item $\ddagger$ denotes the parameter value is taken from \citep{connolly2016long}.
     \item $\dagger\dagger$ adopted from \citep{Leahy_2024}.

\end{tablenotes}
\centering
    \begin{tabular}{||c|c|c|c||}
    \hline
    \hline
    State & Parameters & Values & $\chi^2$(dof)\\
    \hline
    &&&\\
        \multirow{9}{*}{LS1}  & Plasma   & & \\
        &  temperature  & 1.58$_{-0.05}^{+0.05}$& \\
        &(keV) &&\\
        & n$_H$ * (cm$^{-3}$) & 1 $\dagger\dagger$ &\\
        & Abundance * & 1  &\\
          & Redshift * & 0.00428 & 92.49(91)\\
             & Switch * & 2 $\dagger$ &\\
             & N$_H$ * (cm$^{-2}$) & 0.45$\times$10$^{-22}$ $\ddagger$ &\\
         & Spectral Index & 1.92$_{-0.76}^{+0.76}$ & \\
         &&&\\
         \hline
         &&&\\
    \multirow{9}{*}{LS2}  & Plasma  & & \\
     &  temperature & 1.42$_{-0.03}^{+0.03}$& \\
     &(keV)&&\\
      & n$_H$ * (cm$^{-3 }$) & 1 $\dagger\dagger$ &\\
         & Abundance * & 1  &\\
          & Redshift * & 0.00428  &\\
             & Switch * & 2 $\dagger$ & 155.38(119)\\
             & N$_H$ * (cm$^{-2}$) & 0.45$\times$10$^{-22}$ $\ddagger$ &\\
         & Spectral Index & 2.94$_{-0.32}^{+0.32}$ & \\
         &&&\\
         \hline
    \end{tabular}
\end{table}

\end{center}

\begin{table*}
\caption{Flux calculation for Pre-Flare (LS1) and Post-Flare (LS2) in different Swift-UVOT bands.}
\label{Tab:uvot_ls1_ls2_shuffled}
\hspace*{-1.5cm}
\begin{tabular}{||c c c c c c c c||}
\hline
Band & $\lambda_{\mathrm{eff}}$ [\AA] & State & $M_{\mathrm{obs}}$ & $A_\lambda$ & $M_{\mathrm{in}}$ & Zero Point & Flux [$10^{-14}~\mathrm{erg\,cm^{-2}\,s^{-1}\,\AA^{-1}}$] \\
\hline\hline
\multirow{2}{*}{v}    & \multirow{2}{*}{5468} & LS1 & $13.15 \pm 0.02$ & \multirow{2}{*}{0.06} & $13.09 \pm 0.02$ & \multirow{2}{*}{17.88$\pm$0.01} & $40.94 \pm 0.92$ \\
                      &                        & LS2 & $13.15 \pm 0.02$ &                        & $13.09 \pm 0.03$ &                        & $40.94 \pm 1.13$ \\
\hline
\multirow{2}{*}{b}    & \multirow{2}{*}{4392} & LS1 & $14.13 \pm 0.02$ & \multirow{2}{*}{0.08} & $14.05 \pm 0.04$ & \multirow{2}{*}{18.98$\pm$0.02} & $6.14 \pm 0.20$ \\
                      &                        & LS2 & $14.14 \pm 0.02$ &                        & $14.06 \pm 0.04$ &                        & $6.08 \pm 0.19$ \\
\hline
\multirow{2}{*}{u}    & \multirow{2}{*}{3465} & LS1 & $15.26 \pm 0.02$ & \multirow{2}{*}{0.10} & $15.16 \pm 0.04$ & \multirow{2}{*}{19.36$\pm$0.02} & $1.55 \pm 0.04$ \\
                      &                        & LS2 & $15.35 \pm 0.02$ &                        & $15.25 \pm 0.04$ &                        & $1.43 \pm 0.05$ \\
\hline
\multirow{2}{*}{uvw1} & \multirow{2}{*}{2600} & LS1 & $14.98 \pm 0.02$ & \multirow{2}{*}{0.13} & $14.85 \pm 0.05$ & \multirow{2}{*}{18.95$\pm$0.03} & $3.02 \pm 0.13$ \\
                      &                        & LS2 & $16.52 \pm 0.02$ &                        & $16.39 \pm 0.05$ &                        & $0.73 \pm 0.03$ \\
\hline
\multirow{2}{*}{uvm2} & \multirow{2}{*}{2246} & LS1 & $17.14 \pm 0.03$ & \multirow{2}{*}{0.18} & $16.96 \pm 0.05$ & \multirow{2}{*}{18.54$\pm$0.03} & $0.63 \pm 0.03$ \\
                      &                        & LS2 & $15.44 \pm 0.02$ &                        & $15.26 \pm 0.05$ &                        & $3.02 \pm 0.13$ \\
\hline
\multirow{2}{*}{uvw2} & \multirow{2}{*}{1928} & LS1 & $17.06 \pm 0.02$ & \multirow{2}{*}{0.16} & $16.90 \pm 0.05$ & \multirow{2}{*}{19.11$\pm$0.03} & $0.39 \pm 0.02$ \\
                      &                        & LS2 & $17.06 \pm 0.02$ &                        & $16.90 \pm 0.05$ &                        & $0.39 \pm 0.02$ \\
\hline
\end{tabular}
\end{table*}

 We followed the standard data reduction procedure\footnote{\url{https://www.swift.ac.uk/analysis/index.php}} to prepare the raw data, i.e., Level 1 data for analysis, using the software package, HEASoft\footnote{\url{https://heasarc.gsfc.nasa.gov/docs/software/heasoft/}}. \verb|Xrtpipeline| (v0.13.5) has been used to generate the cleaned event files corresponding to each observation. During the analysis, the calibration file (CALDB; version 20240522) and other standard screening criteria have been applied to the cleaned data. We have used the photon count (PC) mode data for our analysis. For non-piled-up spectra, we consider a circular source region between 32$^{\prime\prime}$-- 40$^{\prime\prime}$, with the source at the center. In case of the piled-up spectra, we carried out the standard procedure for pile-up correction\footnote{\url{https://www.swift.ac.uk/analysis/xrt/pileup.php}} and an annular source region centred on the source has been considered. In both cases, the background was extracted from a circular region located far from the source of interest, in an area with minimal contamination from other sources, and with a radius twice that of the source region. Using the \texttt{xselect} tool, we defined the source and background regions and extracted the corresponding spectrum files for each observation. \texttt{xrtmkarf} and \texttt{grppha} programs were utilised to create an ancillary response file and group the spectrum file with its corresponding response matrix file. Then, \texttt{Addspec} and \texttt{mathpha} were employed.
 
In this work, we produced the combined source and background spectra separately for each of the two low states, LS1 and LS2. As mentioned earlier, the `Flare' state does not have any simultaneous Swift observation.  LS1 comprises 11 observations acquired between March 19, 2021, and Jan 11, 2022, and LS2 consists of 26 observations from March 12, 2022, to April 23, 2023.

In the next step, we again used \verb|grppha| to group the combined source and background files of each state and generate the final spectrum file, which was then used in \verb|xspec| for further analysis.
\par
The spectra of the two low states were modelled using the \verb|xspec| package (\cite{xspec_1996A}; Version 12.14.1). We found that both Swift-XRT low state spectra are best described by the \texttt{mekal\footnote{\url{https://heasarc.gsfc.nasa.gov/docs/software/xspec/manual/node199.html}} + (TBabs\footnote{\url{https://heasarc.gsfc.nasa.gov/docs/software/xspec/manual/node283.html}}*powerlaw\footnote{\url{https://heasarc.gsfc.nasa.gov/docs/software/xspec/manual/node222.html}})} model \citep{connolly2016long}, providing a better fit than mekal + (TBabs*logpar\footnote{\url{https://heasarc.gsfc.nasa.gov/docs/software/xspec/manual/node196.html}} ) or mekal + (TBabs*plec\footnote{\url{https://heasarc.gsfc.nasa.gov/docs/software/xspec/manual/node163.html}}). The best-fit parameters were derived using chi-square statistics. Table~\ref{tab:Swift_XRT_Fitting_Parameter} summarises the free and fixed parameter values of the best-fit model. Other than these parameters, the normalisation factors for both the mekal and powerlaw components are left free. Finally, we extracted the unabsorbed spectra corresponding to LS1 and LS2, which were then used in constructing the multi-wavelength spectral energy distributions (MWSEDs) of M87 for these two states.



M87 was also monitored by the Swift Ultraviolet/Optical Telescope (UVOT; \cite{roming2005swift}) with six filters: v (5468~\AA), b (4392~\AA), u (3465~\AA), uvw1 (2600~\AA), uvm2 (2246~\AA), and uvw2 (1928~\AA), at the same time with Swift-XRT. In LS1, the number of observations is 10, 8, and 11 for the v, b, and u bands, respectively, and 10, 11, and 10 for uvw1, uvm2, and uvw2. For LS2, the optical bands have 8 (v), 9 (b), and 16 (u) observations, while the UV bands include 16 (uvw1), 17 (uvm2), and 16 (uvw2) observations.
For the extraction of the source region from the Swift-UVOT data, a circular region of radius 5$^{\prime\prime}$
 was considered, and for the background region, a circle of radius 15$^{\prime\prime}$
 was taken, located far away from the source. We have summed all exposures corresponding to a particular energy band using \verb|uvotimsum|. Then, the `uvotsource' tool has been used to extract the magnitude of the source. This magnitude has been corrected to account for galactic absorption. To obtain the extinction values (\( A_\lambda \)) for each Swift-UVOT filter, the Python package \verb|extinction|\footnote{\url{https://extinction.readthedocs.io/en/latest/}} was utilised.  We have considered the \cite{fitzpatrick1999correcting} dust extinction function with $R_V = 3.1$, where $R_V$ is the slope of the extinction curve. For the diffuse interstellar medium (ISM), this corresponds to a visual extinction $A_V = 0.0612$~mag and reddening $E(B-V) = 0.0197$~mag \citep{schlafly2011measuring}, consistent with the average ISM value of $R_V = 3.1$ \citep{Schultz_1975, Whittet1980,Rieke1985}.
The intrinsic magnitude (\( M_{\text{in}} \)), the extinction parameter (\( A_{\lambda} \)) and the observed magnitude of the source (\( M_{\text{obs}} \)) are related as follows  \(M_{\text{in}}= M_{\text{obs}}-A_{\lambda}\). We have calculated the intrinsic source flux using this relation \( M_{\text{in}} = -2.5 \log_{10}(\text{Flux}) - \mathrm{Zero~point} \). 
The zero point is a calibration parameter that defines the reference count rate corresponding to a standard X-ray flux for each detector mode and energy band in Swift XRT. It allows conversion of raw photon count rates into physically calibrated fluxes or luminosities. Zero points are empirically determined from observations of X-ray standard source with well-characterized spectra.
And finally, we calculated the Swift-UVOT SED of LS1 \& LS2 and used it to construct the MWSED of low state. 
 In Table~\ref{Tab:uvot_ls1_ls2_shuffled}, we have reported the fluxes for the different optical–UV bands from Swift observations of M87, which also lists the band name, wavelength, $M_\mathrm{in}$, $A_\lambda$, $M_\mathrm{in}$, and Zero-point. The Zero-point values are taken from \cite{ZPT_2011}. 

\begin{table}
\centering
\caption{Non-Simultaneous observation times for M87 using different observing facilities spanning various energy bands. \label{tab:NON_SIM_SSDC_DATA_Table}}
\begin{tabular}{||c|c|c||}
\hline
\textbf{Instrument} & \textbf{Start Time} & \textbf{End Time} \\
\hline
Very Large Array  &  &  \\
(VLA; \cite{nagar2001evidence}) & 1999  & - \\
Avg-VLA \citep{prieto} & 2003-06 & 2004-12-31 \\
\tableline
Atacama Large Millimeter/ &  &  \\
submillimeter Array & 2012-06-03 & - \\
 (ALMA; \cite{prieto})&  &  \\
\tableline
 Keck I telescope  &  &  \\
(Long Wavelength  & 2000-01-18 & - \\
 Spectrometer) &  &  \\
 \citep{why} &  &  \\
\tableline
Gemini \citep{perl} & 2001-05 & - \\
\tableline
Hubble Space Telescope & 1998-01-16 & 2003-11-29 \\
(HST \cite{prieto})&& \\
\tableline
MAGIC & 2012 & 2015 \\
\citep{2020MNRAS.492.5354M} &  &  \\
\tableline
Veritas & 2007-02 & 2007-04\\
\citep{Acciari_2008} & 2010-04-05 & 2010-04-15\\
\citep{aliu2012veritas} &&\\
\tableline
HESS (low state  & 2004-02-16 & 2005-02-25 \\
 data in HESS   & 2008-03-11 & 2009-03-21 \\
collaboration 2023)  & 2012-10-31 & 2018-10-15 \\
\citep{aharonian2023constraining} & & \\
\hline
\end{tabular}

\end{table}

\subsection{Archival Data} \label{subsec:Archival_Data}
In this work, we have used the archival data from  Large High Altitude Air Shower Observatory (LHAASO), Monitoring of Jets in Active Galactic Nuclei with VLBA Experiments (MOJAVE), Markarian Multiwavelength Data (MMDC), and Space Science Data Centre (SSDC).

\subsubsection{LHAASO Data}
The LHAASO \citep{LHAASO_Cao:2010zz} is a next-generation ground-based facility in Daocheng, China, designed to investigate the origin of cosmic rays and the mechanisms of VHE gamma-ray production over an extensive energy range from $\sim$100 GeV to 1 EeV. It consists of three major detector systems: the Water Cherenkov Detector Array (WCDA; \cite{LHAASO_WCDA_Li:2014wna}), optimised for gamma-ray observations in the 100 GeV–30 TeV range; the Kilometer Square Array (KM2A), operating in the 30 TeV–1 PeV range for precise cosmic-ray composition studies \citep{KM2A_Sim_Cui:2014bda, KM2A_Aharonian:2020iou}; and the Wide Field-of-View Cherenkov Telescope Array (WFCTA; \cite{LHAASO_WFCTA:2020vgx}), for cosmic ray physics from 10 TeV to 1 EeV. We have used the simultaneous gamma-ray data from the low state and flare state, collected by LHAASO-WCDA (March 8, 2021, to March 16, 2024) and LHAASO-KM2A (July 20, 2021, to January 31, 2024) from \cite{LHAASO_M87:2024}.

\subsubsection{MOJAVE Data}
The MOJAVE is a long-term program to monitor radio brightness and polarisation in AGN jets across the northern sky at 15 GHZ \citep{MOJAVEWebpage}. With an angular resolution of greater than 1 mas, MOJAVE captures a full polarisation image by monitoring at these wavelengths: 7 mm, 1.3 cm, and 2 cm. We utilised simultaneous data from MOJAVE \footnote{\url{https://www.cv.nrao.edu/MOJAVE/sourcepages/1228+126.shtml}}. We have used one data set for LS1, collected on May 1, 2021, and four data sets for LS2, spanning from March 18, 2022, to January 28, 2024, from the MOJAVE/2 cm Survey Data Archive to build the MWSED. No simultaneous MOJAVE data has been found that corresponds to the Flare state.

\subsubsection{MMDC Data}
The MMDC\footnote{\url{https://mmdc.am/}} (\cite{MMDC_1_2024, MMDC_3_2024, MMDC_2_2024}) is a web-based tool designed to build and analyse multiwavelength data from blazars. In our work, we have collected multi-wavelength data from MMDC, which includes $i)$ optical data from the All-Sky Automated Survey for Supernovae (ASAS-SN) and $ii)$ infrared data from the Near-Earth Object Wide-field Infrared Survey Explorer (NEOWISE) and the Zwicky Transient Facility (ZTF) for both LS1 \& LS2. For the flare state, we have only found simultaneous optical data collected from ASAS-SN.

\subsubsection{SSDC Data}
\begin{figure*}
    \centering
    \includegraphics[width=0.9\linewidth]
    {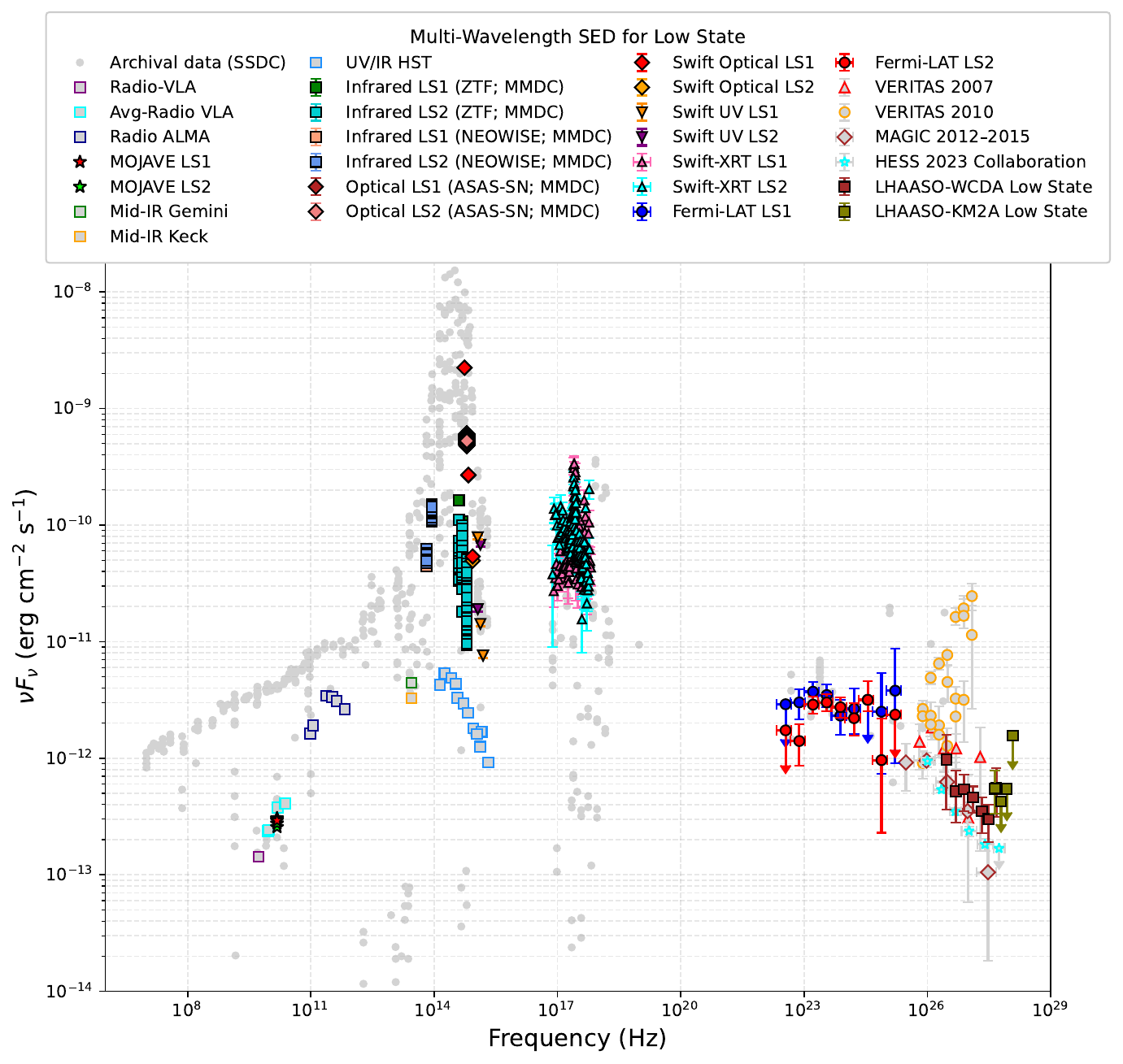}
    
    \caption{Multi-wavelength SED plot for low state. Non-simultaneous data points are represented by markers with solid grey \& grey-fill coloured edges, whereas simultaneous data points are represented by markers filled with one colour and outlined with black colour.}
    \label{fig:lowstate_mwsed}
\end{figure*}

\begin{figure*}
    \centering
    \includegraphics[width=0.9\linewidth]
    {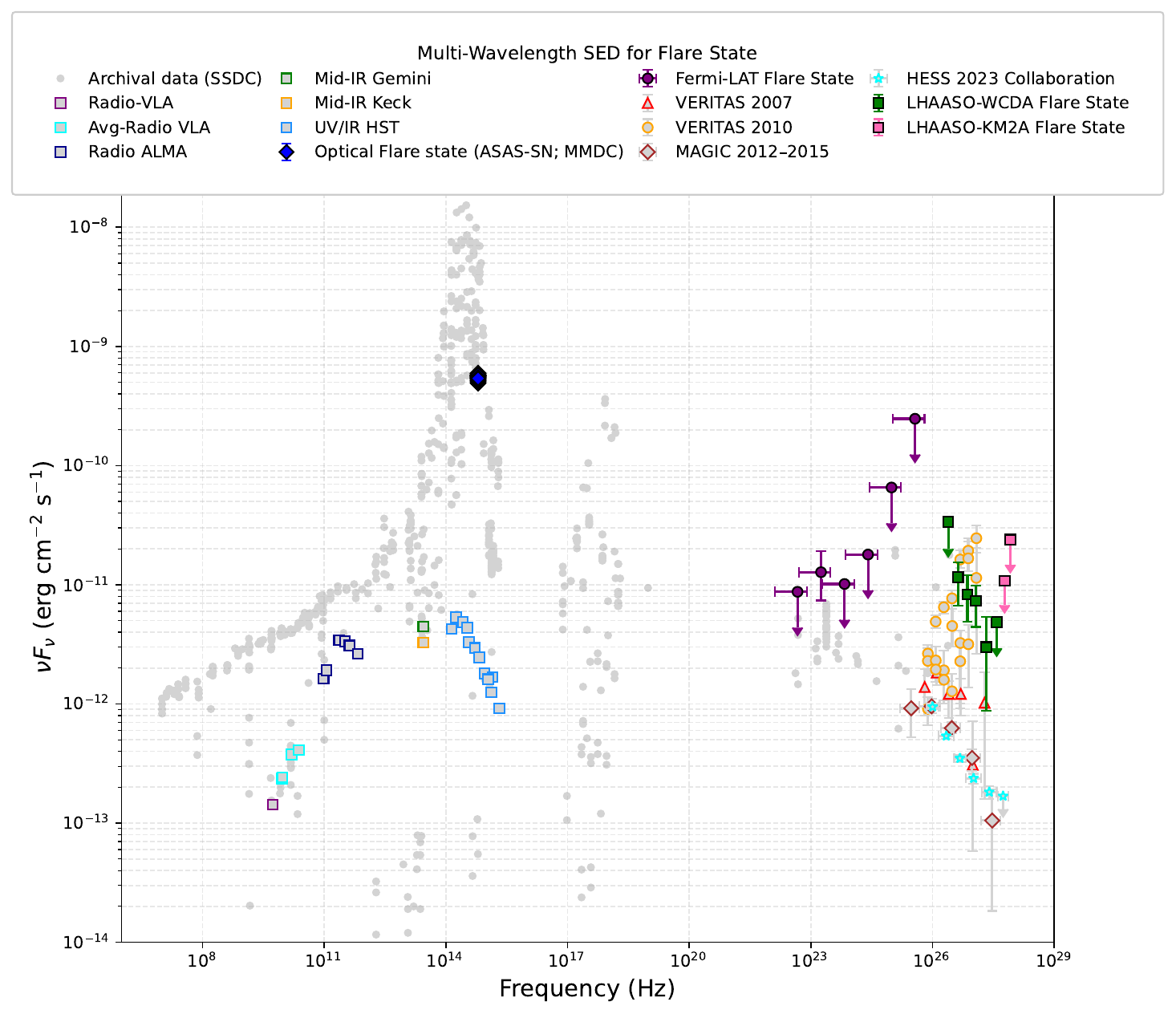}
    
    \caption{Multi-wavelength SED plot for flare state.
    The marker style and colour conventions are similar to Fig.~\ref{fig:lowstate_mwsed}.}
    \label{fig:flare_mwsed}
\end{figure*}

We gathered multi-wavelength data from SSDC SED Builder\footnote{\url{https://tools.ssdc.asi.it/SED/}}, which is displayed in \autoref{fig:lowstate_mwsed},  \autoref{fig:flare_mwsed}, \autoref{Fig:Low-state_GAMERA} and \autoref{Fig:Flare-state_GAMERA} as grey circles.

\subsubsection{Other Data Sources}
We collected non-simultaneous data from various observational facilities, listed in Table~\ref{tab:NON_SIM_SSDC_DATA_Table}.


\section{Multi-wavelength SED modelling:}
\label{sec:MWSED_modelling}

We constructed the MWSED by compiling the analysed \textit{Fermi}-LAT, Swift (see sec :~\ref{subsubsec:Fermi_LAT_Data_Analysis} sec:~\ref{subsubsec:Swiftxrt_UVOTT_Data_Analysis}) and the archived data (see sec :~\ref{subsec:Archival_Data}), covering radio to gamma-ray frequencies. The MWSEDs corresponding to the low (combining LS1 \& LS2) and flare states are shown in Fig.~\ref{fig:lowstate_mwsed} and Fig.~\ref{fig:flare_mwsed}, respectively.

We consider a homogeneous spherical emission region of radius $R'$ having a uniform magnetic field $B'$ inside it, moving through a relativistic jet of bulk Lorentz factor $\Gamma_b$ (corresponding to Doppler factor $\delta$). This region contains relativistic electrons, which lose energy via synchrotron radiation and inverse-Compton (IC) processes. We are considering a leptonic model in which the energy loss of protons is insignificant compared to that of electrons.
\par
As discussed in Sec :~\ref{sec:Fermi-LAT_mwsed}, from the maximum likelihood analysis of \textit{Fermi}-LAT data, a log-parabolic distribution of gamma-rays was found to best fit the gamma-ray data of the flare and low states. A log-parabolic photon spectrum can be produced from the radiative loss of a log-parabolic electron spectrum \citep{massaro}. Therefore, we have used a log-parabolic spectrum for the injected electrons in the emission region to explain the MWSED of M87, given by the following expression:

\begin{equation} 
Q(E') = L_0 \left(\frac{E'}{E'_0}\right)^{-\alpha - \beta \log_{10}\left(\frac{E'}{E'_0}\right)},
\end{equation}

where $Q(E')$ is the log-parabolic distribution, $L_0$ is the normalisation constant, $E'_0$ is the scaling factor or pivot energy (set to 1 TeV in our modelling and kept fixed), $\alpha$ is the spectral index, and $\beta$ is the curvature index.

\par
We use ‘GAMERA’ \citep{GAMERA_Hahn:2015hhw}, an open-source code to model the MWSEDs of M87, available on GitHub\footnote{\url{http://libgamera.github.io/GAMERA/docs/main_page.html}}. It solves the time-dependent transport equation (Eqn.~\ref{eqn:GAMERA_Transport_Eqn}) and propagates the particle spectrum $N(E^\prime, t^\prime)$
 for an injected spectrum to calculate the synchrotron and IC emissions, including the Klein–
Nishina effect. 
\begin{align} 
\frac{\partial N(E',t')}{\partial t'} &= Q(E',t') - \frac{\partial}{\partial E'}\left(b(E',t')N(E',t')\right) \nonumber \\
&\quad - \frac{N(E',t')}{\tau'_{\text{esc}}(E',t')} \label{eqn:GAMERA_Transport_Eqn}
\end{align}

$Q(E^\prime, t^\prime)$
 is the injected electron spectrum and $b(E^\prime, t^\prime)$
 corresponds to the energy loss rate by synchrotron and IC, and can be defined as \big($\frac{dE'}{dt'}$\big). In the last term, the escape time scale is assumed to be independent of energy and time, $\tau'\textsubscript{esc} = \eta'_{esc}\frac{R'}{c}$, where $\eta'_{esc}$ denotes the coefficient of escape time. 
\par
The code, subsequently, calculates the synchrotron and synchrotron self-Compton (SSC) emissions, which are Doppler boosted by a factor of $\delta^4$ \citep{2009herb.book.....D} in the observer's frame due to relativistic beaming, where, $\delta = [\Gamma_b(1-\beta_j cos\theta)]^{-1}$ is the Doppler factor, $\Gamma_b$ is the bulk Lorentz factor, $\beta_j$ is the intrinsic speed of the jet frame and $\theta$ is the viewing angle of the jet with respect to the line of sight of the observer. 
\par
We have assumed two emission regions, one in the sub-parsec scale jet and another in the kilo-parsec scale jet of M87. The light curve of the VHE flare data recorded by the LHAASO collaboration \citep{LHAASO_M87:2024} suggests a minimum variability timescale of one day, which constrains the size of the emission region. In our model, the VHE flare happened 
from the sub-parsec scale jet; however, the kiloparsec scale jet gives a dominant contribution to VHE gamma-ray data in the low state.
In the sub-parsec scale jet, the gamma-rays are produced by the SSC mechanism, whereas in the kilo-parsec scale jet, they are mostly emitted by the external Compton mechanism (EC). In the EC mechanism, the external radiation fields provide the target photons for the inverse Compton scattering. The external radiation fields, e.g., CMB, starlight, and the dust emitted photons, are isotropic with respect to the AGN frame. The starlight and dust energy density of the host galaxy has been used from \citet{2003ApJ...597..186S} to calculate the EC emission from the kilo-parsec jet of M87. The high-energy gamma-rays produced in external Compton emission are Doppler-boosted by the factor of $\delta^6$/$\Gamma^2$ in the observer's frame \citep{2009herb.book.....D}. 
\par
We have assumed the values of the physical parameters of the jets, for example, the magnetic field \citep{tomar2021broadband} and the Lorentz factor (\citet{Walker_2018},\citet{2009MNRAS.395..301W}), following the earlier studies on the jets of M87.

\par
The parameters listed in the Table~\ref{tab:GAMERA_Lowstate_modeling} and Table~\ref{tab:GAMERA_Flare_modeling} are adjusted to fit the simulated spectral energy distributions to the data points.
\par
We used the following relation to determine the kinematic jet power of the ﬂaring and 
 low states in sub-parsec and kilo-parsec scale jet, tabulated in 
 Table~\ref{tab:Jet_Power} :\begin{equation}
    P_{tot} = \pi R'^2 \Gamma_{b}^2 c (U\ensuremath{'}_{e} + U\ensuremath{'}_B + U\ensuremath{'}_p)
\end{equation}
The quantities in the AGN frame and the comoving jet frame are indicated by the unprimed and primed notations, respectively. $U\ensuremath{'}_{e}, U\ensuremath{'}_{B}$ and $U\ensuremath{'}_{p}$  are the energy densities of electrons, magnetic field, and protons in the comoving jet frame. These are defined as follows:

\begin{equation}
    U'_e = \frac{1}{V'} \int_{E'_{\min}}^{E'_{\max}} Q(E')\,E'\, dE'
\end{equation}, V is the volume of the emission region.

\begin{equation}
    U\ensuremath{'}_B = \frac{B'^2}{8\pi}
\end{equation}

\begin{equation}
    U\ensuremath{'}_p =n_p\, m_p\,c^2 
\end{equation}, $m_p$ is the mass of proton.
We assumed that the no of protons ($n_p$) and radiating electrons is equal in the jet to maintain charge neutrality.
\par
 According to \citet{2019ApJ...875L...6E}, 
 the estimated mass of the black hole ($M_{BH}$) is \( (6.5 \pm 0.7) \times 10^{9}\, M_{\odot} \)
for an assumed distance of 16.8 Mpc, $M_\odot$ is the solar mass. Using this value, we calculated the Eddington luminosity $L_{\mathrm{Edd}} \approx 8.19 \times 10^{47}\ \mathrm{erg\ s^{-1}}$. The total kinematic jet power obtained in our model is less than the Eddington luminosity of M87.

\begin{figure*}[htbp] 
    \centering
    \includegraphics[width=0.9\linewidth]{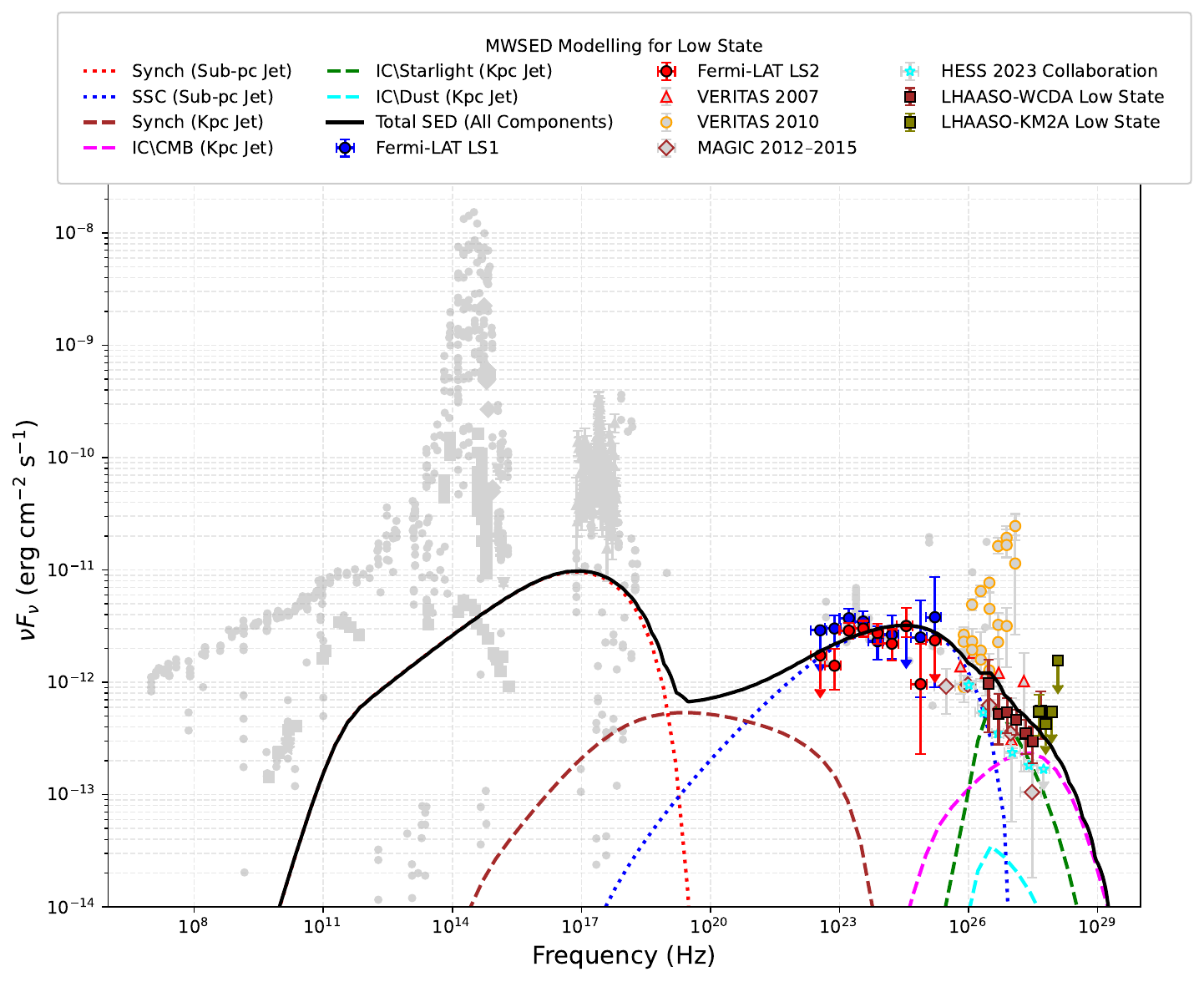}
    \caption{MW modelling of the low state of M87 with contributions from the sub-parsec and kiloparsec-scale jets.
    SSDC data and host galaxy emission are shown in solid grey colour.
    The red and blue dotted lines represent the synchrotron and SSC emission from the sub-pc jet, respectively. The brown, magenta, green, and cyan dashed lines indicate the synchrotron, IC/CMB, IC/Starlight, and IC/Dust components in kpc-jet. The sum of all of these components (total SED) is shown by the black solid line.}
    \label{Fig:Low-state_GAMERA}
\end{figure*}
\begin{figure*}[htbp] 
    \centering
    \includegraphics[width=0.9\linewidth]{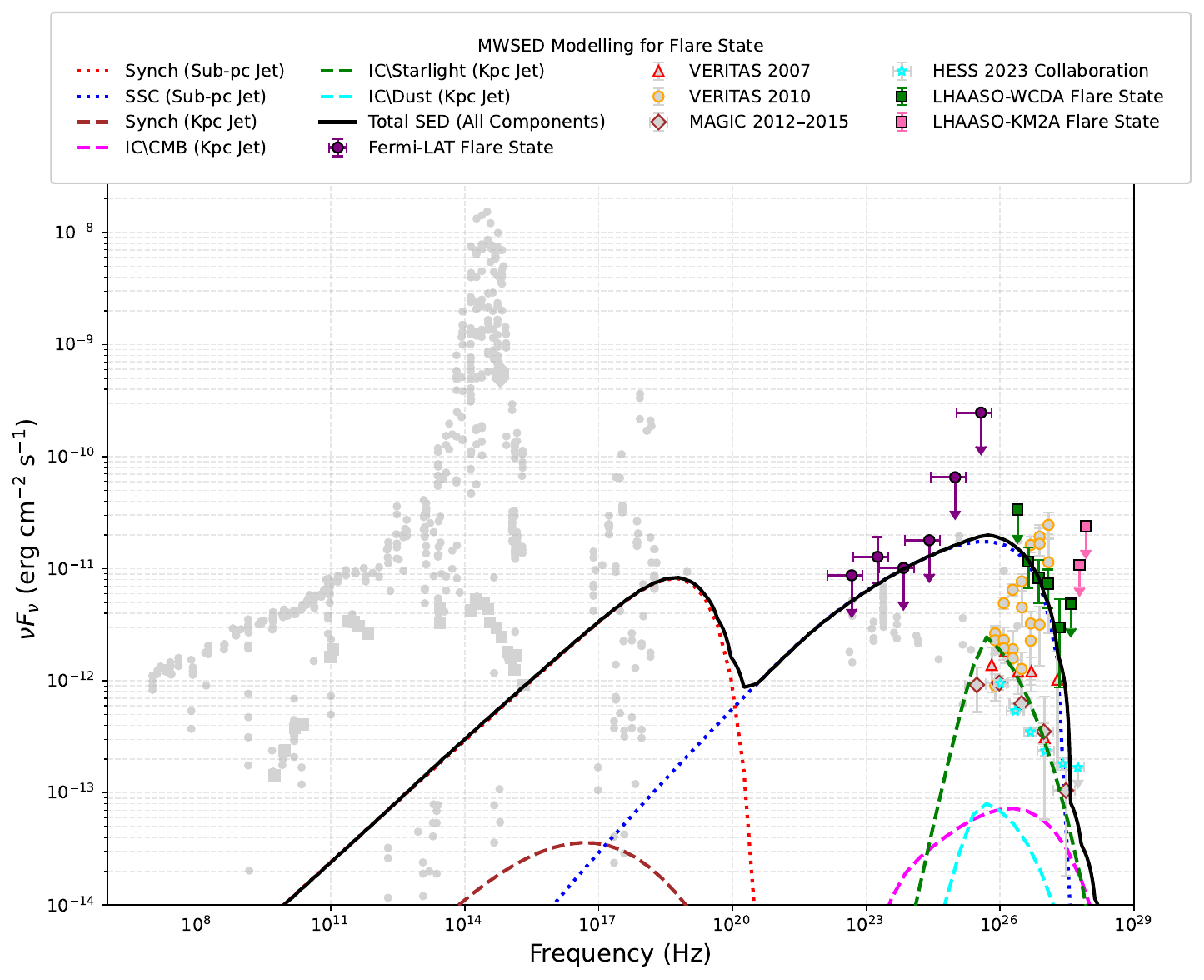}
    \caption{MW modelling of the flare state of M87 with contributions from the sub-parsec and kiloparsec-scale jets. The line styles are similar to Fig.~\ref{Fig:Low-state_GAMERA}. SSDC data and host galaxy emission are shown in solid grey colour.}
    \label{Fig:Flare-state_GAMERA}
\end{figure*} %


\section {Result: Multi-wavelength Modelling} \label{sec:result}

The LS1 and LS2 states were combined and labelled as the low state in this work. We modelled the MWSED of this state, with the SSC emission from the sub-parsec scale jet. This model successfully explains the \textit{Fermi}-LAT MeV–GeV observations as the \textit{Fermi}-LAT data points are covered by the SSC bump. However, the SSC emission from the sub-parsec scale jet is inadequate to account for the LHAASO-WCDA and LHAASO-KM2A data at higher energies (Fig.~\ref{Fig:Low-state_GAMERA}), hence, we needed another emission region to explain the VHE gamma-ray emission. We considered the contribution from the kiloparsec-scale jet of M87 for this purpose. At the kiloparsec-scale jet, the EC emission is significantly higher compared to the SSC emission, as the emission region has a radius of the order of kilo-parsec ($\sim 10^{21}$ cm) \citep{2020Nature}. The external photon fields, such as starlight, thermal dust emission, and the cosmic microwave background (CMB), acted as targets for IC scattering. The corresponding energy densities of the external photon fields were adopted from \citet{2003ApJ...597..186S}. The total emission from the kiloparsec-scale jet consists of the following components: synchrotron, SSC, IC/Starlight, IC/CMB, and IC/Dust. As shown in Fig.~\ref{Fig:Low-state_GAMERA}, IC/Starlight dominates and largely accounts for the observed TeV flux; the contribution from IC/CMB is lower than the IC/Starlight; and the contribution from IC/Dust is lower than the IC/CMB. The solid black curve in Fig.\ref{Fig:Low-state_GAMERA} represents the total SED combining the sub-parsec and kiloparsec jets. The hardening of the VHE gamma-ray spectrum has been attributed to the EC emission from the kiloparsec-scale jet. The solid black curve passes through the first LHAASO-WCDA data point and the upper error bars of the subsequent LHAASO-WCDA data points. It also passes through the lower error bars of the LHAASO-WCDA and LHAASO-KM2A data points at the highest energy. 
Thus, our model is successful in explaining the complete gamma-ray spectrum covering the \textit{Fermi}-LAT and LHAASO energy ranges. The model parameters are summarised in 
Table~\ref {tab:GAMERA_Lowstate_modeling}. The total resulting jet power, reported in Table~\ref{tab:Jet_Power}, remained well below the Eddington limit of this source.

\par
We modelled the flare-state MWSED with the SSC emission from the sub-parsec jet, as the light curve analysis by the LHAASO collaboration already constrained the size of the emission region during the flare state. Fig.~\ref{Fig:Flare-state_GAMERA} shows that the SSC model can explain the simultaneous GeV–TeV gamma-ray data. The solid black line, representing the total SED, touches the lower error bar of the \textit{Fermi}-LAT data point, and it is below the \textit{Fermi}-LAT upper limits. The LHAASO-WCDA data points shown with green coloured boxes are well covered by the solid black line. The contribution from the kiloparsec-scale jet was also included, although it remained subdominant compared to the sub-parsec jet emission during the flare. The fitted parameters are provided in Table~\ref{tab:GAMERA_Flare_modeling}, and the powers of the individual components, along with the total power of the jet, are listed in Table~\ref{tab:Jet_Power}. In this case also, the total jet power remained several orders of magnitude below the Eddington luminosity.
\par
The low-energy data show sharp peaks which are produced by host galaxy emission (\citet{lucchini}, \citet{2025arXiv250801986X}). We have not fitted these peaks, as this work focuses on the jet emission and explains the origin of the very high-energy gamma-rays detected from M87.
\begin{table*}[htbp]
\caption{Results of multi-wavelength SED modelling for sub-parsec and kilo-parsec scale jets in the low state, corresponding to \autoref{Fig:Low-state_GAMERA}.}
\centering
\begin{tabularx}{\textwidth}{l c c c c }   
\hline\hline
Parameter & Symbol & Unit & Sub-parsec jet & Kilo-parsec jet \\
\hline
Spectral index of injected electron spectrum & $\alpha$ & -- & 2.5 & 2.0 \\
Curvature index of injected electron spectrum & $\beta$ & -- & 0.019 & 0.021 \\
Escape time coefficient & $\eta'_{esc}$ & -- & 1 & 1 \\
Magnetic field in emission region & $B'$ & G & 0.04 & $4\times10^{-6}$ \\
Size of the emission region & $R'$ & cm & $1.8\times10^{16}$ & $1.8\times10^{21}$ \\
Viewing angle & $\theta$ & deg & $17^\circ{}^a$ & $15^\circ{}^b$ \\
Doppler factor & $\delta$ & -- & 3.3$^a$ & 2.29$^b$ \\
Lorentz factor & $\Gamma_b$ & -- & 2.7$^a$ & 1.45$^b$ \\
Min. Lorentz factor of injected electrons & $\gamma'_{min}$ & -- & $5.65\times10^{2}$ & $1\times10^{7}$ \\
Max. Lorentz factor of injected electrons & $\gamma'_{max}$ & -- & $3.5\times10^{6}$ & $3\times10^{11}$ \\
Star light energy density & $U'_{star}$ & erg cm$^{-3}$ & -- & $2.10\times10^{-9}$ \\
Dust energy density & $U'_{dust}$ & erg cm$^{-3}$ & -- & $2.10\times10^{-11}$ \\
CMB energy density & $U'_{CMB}$ & erg cm$^{-3}$ & -- & $8.76\times10^{-13}$ \\
\hline
\hline
\end{tabularx}
\begin{tablenotes}
      \small
      \item a adopted from \citet{Walker_2018}
      \item b adopted from \citet{2009MNRAS.395..301W}
\end{tablenotes}
\label{tab:GAMERA_Lowstate_modeling}
\end{table*}

\begin{table*}[htbp]
\caption{Results of multi-wavelength SED modelling for sub-parsec and kilo-parsec scale jets in the flare state corresponding to \autoref{Fig:Flare-state_GAMERA}.}
\centering
\begin{tabularx}{\textwidth}{l c c c c}   
\hline\hline
Parameter & Symbol & Unit & Sub-parsec jet & Kilo-parsec jet \\
\hline
Spectral index of injected electron spectrum & $\alpha$ & -- & 2.29 & 2.21 \\
Curvature index of injected electron spectrum & $\beta$ & -- & 0.0019 & 0.17 \\
Escape time coefficient & $\eta'_{esc}$ & -- & 1 & 1 \\
Magnetic field in emission region & $B'$ & G & 0.04 & $4\times10^{-6}$ \\
Size of the emission region & $R'$ & cm & $1.5\times10^{15}$ & $1.8\times10^{21}$ \\
Viewing angle & $\theta$ & deg & $17^a$ & $15^b$ \\
Doppler factor & $\delta$ & -- & 3.3$^a$ & 2.29$^b$ \\
Lorentz factor & $\Gamma_b$ & -- & 2.7$^a$ & 1.45$^b$ \\
Min. Lorentz factor of injected electrons & $\gamma'_{min}$ & -- & 40 & $1.65\times10^6$ \\
Max. Lorentz factor of injected electrons & $\gamma'_{max}$ & -- & $10.5\times10^6$ & $2.29\times10^{11}$ \\
Star light energy density & $U'_{star}$ & erg cm$^{-3}$ & -- & $2.10\times10^{-9}$ \\
Dust energy density & $U'_{dust}$ & erg cm$^{-3}$ & -- & $2.10\times10^{-11}$ \\
CMB energy density & $U'_{CMB}$ & erg cm$^{-3}$ & -- & $8.76\times10^{-13}$ \\
\hline
\hline

\end{tabularx}
\begin{tablenotes}
      \small
      \item a adopted from \citet{Walker_2018}
      \item b adopted from \citet{2009MNRAS.395..301W}
\end{tablenotes}
\label{tab:GAMERA_Flare_modeling}
\end{table*}
\begin{center}
    \begin{table*}[htbp]
    \centering
     \caption{Estimated jet power of individual components and total jet for different states}
        
    \label{tab:Jet_Power}
    
\centering
    \begin{tabular}{||c|c|c|c|c||}
    \hline
    \hline
    State & Parameters & Sub-Parsec Jet & Kilo-Parsec Jet & Total Jet Power ($P_{tot}^{sub-pc}+P_{tot}^{kpc}$) \\
    \hline
    &&&&\\
    \multirow{3}{*}{Low State}  & Jet power of relativistic electrons ($P_e$)  & $1.60\times10^{43}$ erg s$^{-1}$ & $1.56\times10^{41}$ erg s$^{-1}$& \\
        
        & Jet power of magnetic field ($P_B$) & $1.42\times10^{40}$ erg s$^{-1}$ & $4.09\times10^{41}$ erg s$^{-1}$ & $3.20\times10^{43}$ erg s$^{-1}$\\

        & Jet power of cold protons ($P_p$) & $1.55\times10^{43}$ erg s$^{-1}$ & $4.01\times10^{36}$ erg s$^{-1}$ &\\
        
         & Total kinematic jet power ($P_{tot}$) & $3.14\times10^{43}$ erg s$^{-1}$ & $5.65\times10^{41}$ erg s$^{-1}$&\\

         &&&&\\
         \hline
         &&&&\\
        \multirow{3}{*}{Flare State}  & Jet power of relativistic electrons ($P_e$) & $3.97\times10^{43}$ erg s$^{-1}$ & $1.36\times10^{41}$ erg s$^{-1}$ &\\
        
        & Jet power of magnetic field ($P_B$) & $9.84\times10^{37}$ erg s$^{-1}$ & $4.09\times10^{41}$ erg s$^{-1}$ & $4.51\times10^{44}$ erg s$^{-1}$\\

       & Jet power of cold protons ($P_p$) & $4.11\times10^{44}$ erg s$^{-1}$ & $5.09\times10^{37}$ erg s$^{-1}$& \\
        
         & Total kinematic jet power ($P_{tot}$) & $4.50\times10^{44}$ erg s$^{-1}$ & $5.45\times10^{41}$ erg s$^{-1}$&\\
        
         &&&&\\
         \hline
         \hline
    \end{tabular}
    
\end{table*}

\end{center}
\section{Discussion and Conclusion}
\label{sec:discussion}

Several emission models have been suggested previously to explain the MWSED of the M87 jet.
A proton synchrotron model was discussed in this paper \citet{2004A&A...419...89R}. \citet{2016ApJ...830...81F} discussed photo-hadronic interactions to explain the very high-energy gamma-ray data of M87. In the paper by \citet{2022ApJ...938...79B}, the fitting of the SED involves different emission mechanisms which combine jet emission and accretion flow. The high-energy bump in the gamma-ray band is explained by proton synchrotron emission, and the X-ray energy band is attributed to the Comptonization of advection-dominated accretion flow (ADAF) photons. \citet{lucchini} explained the radio to X-ray emission by a multizone model, which yields a lower gamma-ray flux than the measured flux. Two distinct one-zone leptonic models were employed in \citet{2021ApJ...911L..11E}, but they failed to simultaneously represent the high and low energy components of the SED of M87. \citet{Alfaro_2022} has shown that the one-zone SSC model, along with the photo-hadronic model, can explain the MWSED covering radio to VHE gamma-ray data.  They included H.E.S.S., MAGIC, and HAWC data in their SED modelling. It is evident from their work that a separate emission model is needed to explain the VHE data of the M87 jet. The VHE gamma-ray data observed by the LHAASO collaboration show spectral hardening near 20 TeV, which has been explained by \citet{2025arXiv250801986X}  with the $\pi^{0}$ decay gamma-rays produced in pp interactions, and also with the two-zone SSC model.

\par
\citet{2019A&A...623A...2A} argued that the one-zone SSC model is challenged by a potential excess over the standard power-law model at around 10 GeV during the low-state of M87, which suggests the need for a second radiation component. The X-ray emissions from the knots in the kilo-parsec scale of M87 have been studied earlier by reanalysing the archival data from \textit{Chandra} observations \citep{2018sun}. \citet{2018sun} modelled radio to X-ray data from the knots of the kilo-parsec scale jet of M87  with synchrotron emission of relativistic electrons. They noted that for some knots, a second component is needed to explain the high-energy data. \citet{tomar2021broadband} demonstrated that synchrotron and SSC emission from relativistic electrons in a sub-parsec scale jet can account for the observed multi-wavelength data from M87 up to 8 GeV.  In our work, we used a two-zone leptonic model to fit the broadband SED of M87; the two zones are located at the sub-parsec and the kilo-parsec scale jets, respectively. In the low-state, the \textit{Fermi}-LAT MeV–GeV data points are satisfactorily covered by the SSC emission from the sub-parsec scale jet. LHAASO collaboration measured the emission of VHE gamma-rays with a flux level of \(10^{-12}~\mathrm{erg}~\mathrm{cm}^{-2}~\mathrm{s}^{-1}\) \citep{LHAASO_M87:2024} during this state. It was noted in their work that the low-state VHE emission has a higher possibility of coming from large-scale structures, such as the bright knots HST-1, knot A in the kiloparsec-scale jet. We have shown that the LHAASO data points are well fitted by the VHE gamma-ray spectrum produced in EC emission of the ultra-relativistic electrons accelerated at the kiloparsec-scale jet of M87. This model can explain the hardening in the VHE gamma-ray spectrum near 20 TeV.

\par
LHAASO collaboration \citep{LHAASO_M87:2024} also reported a flare in their VHE gamma-ray data, which lasted for 8 days in 2022. They observed that this flare rose fast ($\tau_{\rm d}^{\rm rise} = 1.05\pm0.49$ days) and decayed at a slower rate ($\tau_{\rm d}^{\rm decay} = 2.17\pm0.58$ days), and mentioned that its dayscale variability indicates that the VHE emission originates from a compact emission region that is only a few Schwarzschild radius from the centre of the supermassive black hole, with a radius of $R\sim 2.7\times10^{15}\delta\, {\rm cm}$ (doppler factor $\delta  {\la 3}$). This implies that the acceleration and emission process of non-thermal particles may occur close to the central black hole. In earlier studies, the underlying physical mechanisms of such flares have been speculated.
According to recent high-resolution general relativistic magnetohydrodynamic simulations, flares close to the black hole event horizon may be powered by episodic magnetic reconnection events (\cite{2020ApJ...900..100R}, \cite{2023ApJ...943L..29H}). Another possible argument (\cite{2010ApJ...724.1517B}, \cite{2017ApJ...841...61A}) is that an efficient production of VHE photons through hadronic processes may result from interactions between a relativistic magnetised outflow near the jet formation zone and a star that was partially tidally disrupted by the contact with the SMBH. In our work, the SSC emission by the relativistic electrons in the sub-parsec scale jet has been used to model the flare-state MWSED of M87 within a time-dependent framework; the GeV-TeV gamma-ray data are well explained by this model. The size of the emission region used in our model is comparable to that mentioned in the paper by the LHAASO collaboration \citep{LHAASO_M87:2024}.

\par
In conclusion, our two-zone leptonic model explains the flare and low state data of LHAASO, the hardening in the LHAASO data in the low state, and the variability time in the flare state. Future observational data from LHAASO-WCDA and LHAASO-KM2A would be useful to understand more about the hardening in the spectrum of M87, as well as the measurement of the variability time from the gamma-ray light curve would be helpful to explore more about the location and size of the emission region.

\section{Software and third-party data repository citations}
The \textit{Fermi}-LAT gamma-ray data analysis was done with “Fermipy” \citep{Fermipy_Version}. Swift X-ray, ultraviolet, and optical data have been analysed with “HEASoft” \citep{Heasoft_2014ascl.soft08004N}. We used data in our work from the MOJAVE database, which the MOJAVE team maintains \citep{MOJAVE_2018}, data from Space Science Data Centre (SSDC) and  Markarian Multiwavelength Data Center (MMDC; \cite{MMDC_1_2024, MMDC_3_2024, MMDC_2_2024}).

\begin{acknowledgments}
The data analysis for this work was performed using the computing facility at the Raman Research Institute. N.M. and S.K.M. thank A.K. Das for valuable discussions on X-ray data analysis. We thank the referee for many helpful comments.
\end{acknowledgments}

\facilities{Fermi-LAT, Swift(XRT and UVOT)}

\software
{Fermipy (V1.0.1; \url{https://fermipy.readthedocs.io/en/latest/} ;\citet{Fermipy_Version}), HEAsoft (V6.26.1; \url{https://heasarc.gsfc.nasa.gov/docs/software/heasoft/}), Xspec package (V12.14.1; \citet{xspec_1996A}), GAMERA (\url{https://github.com/libgamera/GAMERA}; \citet{GAMERA_Hahn:2015hhw}}.


\renewcommand{\thetable}{A\arabic{table}}  
\setcounter{table}{0}                      

\bibliographystyle{aasjournal}	
\bibliography{M87}{}

\end{document}